\documentclass[a4paper,12pt,draftcls,technote,onecolumn]{IEEEtran}%
\usepackage{mathrsfs}
\usepackage{amsfonts}
\usepackage{amsmath}
\usepackage{amssymb}
\usepackage{graphicx}%

\ifCLASSINFOpdf
\else
\fi
\begin{document}
%
\title{Output Feedback Tracking Control for a Class of Uncertain Systems subject to
Unmodeled Dynamics and Delay at Input}
%
%
%

\author{Quan Quan, Hai Lin, Kai-Yuan Cai
\thanks{Quan Quan is with Department of Automatic Control, Beihang University, Beijing, 100191,
China, was with Department of Electrical and Computer Engineering,
National University of Singapore, Singapore, 117576, Singapore. Email: (see http://quanquan.buaa.edu.cn).}
\thanks{Hai Lin is with Department of Electrical and Computer Engineering,
National University of Singapore, Singapore, 117576,
Singapore.}
\thanks{Kai-Yuan Cai  is with Department of Automatic
Control, Beihang University, Beijing, 100191,
China.}
}

%
%

\markboth{}%
{Shell \MakeLowercase{\textit{et al.}}: Bare Demo of IEEEtran.cls for Journals}
%



\maketitle

\begin{abstract}
Besides parametric uncertainties and disturbances, the unmodeled
dynamics and time delay at the input are often present in practical
systems, which cannot be ignored in some cases. This paper aims to
solve output feedback tracking control problem for a class of
nonlinear uncertain systems subject to unmodeled high-frequency
gains and time delay at the input. By the additive state
decomposition, the uncertain system is transformed to an
uncertainty-free system, where the uncertainties, disturbance and
effect of unmodeled dynamics plus time delay are lumped into a new
disturbance at the output. Sequently, additive state decomposition
is used to decompose the transformed system, which simplifies the
tracking controller design. To demonstrate the effectiveness, the
proposed control scheme is applied to three benchmark examples.
\end{abstract}

\begin{IEEEkeywords}
Additive state decomposition, tracking, input delay, unmodeled
dynamics, output feedback, nonlinear systems.
\end{IEEEkeywords}

%
\IEEEpeerreviewmaketitle

\section{Introduction}

Tracking control of an uncertain system is a challenging problem.
Most of research mainly focuses on systems subject to parametric
uncertainties and additive disturbances
\cite{Eduardo(2009)}-\cite{Ferreira(2011)}. Also, some research
focuses on systems subject to uncertainties at the input, such as
backlash, dead zone or other nonlinearities \cite{Corradini(2003)}%
-\cite{Chaturvedi(2006)}. It is well known that unmodeled dynamics
and time delay at the input are also often present in practical
systems. For example, the unmodeled dynamics and time delay at the
input often exist in flight control systems
\cite{Smith(1986)}-\cite{Johnson(1994)}. These uncertainties at the
input may produce a significant degradation in the tracking
performance or even cause instability if not dealt with properly. In
the literature, there are some academic examples to demonstrate that
uncertainties at the input cannot be ignored in some cases. For
example, in \cite{Rohrs(1982)}, the authors constructed a simple
example, later known as \textit{Rohrs' example,} to show that
conventional adaptive control algorithms lose their robustness in
the presence of unmodeled dynamics. Also, some control algorithms
may lose their robustness in the presence of input delay, see for
example the repetitive control example considered in
\cite{Quan(2011)}. Therefore, it is important to explicitly consider
unmodeled dynamics and time delay at the input in the controller
design.

In this paper, the output feedback tracking control problem is
investigated for a class of single-input single-output (SISO)
nonlinear systems subject to mismatching parametric uncertainty,
mismatching additive disturbances, unmodeled high-frequency gains
and time delay at the input. Before introducing our main idea, some
accepted control methods in the literature to handle uncertainties
are briefly reviewed. A nature way is to estimate all of the unknown
parameters, then compensate for them. In \cite{Delphine(2009)}, the
tracking problem for a linear system subject to unknown parameters
and the unknown input delay was considered, where both the
parameters and input delay were estimated by the proposed method.
However, this method cannot handle unparameterized uncertainties
such as unmodeled high-frequency gains. The second way is to design
adaptive control with robustness against unmodeled dynamics and time
delay at the input. In \cite{Enric(2009)}, the Rohrs' example and
the two-cart example, which are tracking problems for uncertain
linear systems subject to unmodeled dynamics and time delay at the
input respectively, were revisited by the $\mathcal{L}_{1}$ adaptive
control. In \cite{Cao(2010)}, the authors analyzed that their
proposed method is robust against time delay at the input. The third
way is to convert a tracking problem to a stabilization problem by
the idea of internal model principle \cite{Francis(1976)}, if
disturbances or desired trajectories are limited to a special case.
In \cite{Trinh(1996)}, the problem of set point output tracking of
an uncertain linear system with multiple delays in both the state
and control vectors was considered. There also exist other methods
to handle uncertainties. However, some of them such as high-gain
feedback cannot be applied to the considered system directly as they
rely on rapid changing control signal to attenuate uncertainties and
disturbance. After passing unmodeled high-frequency gains or time
delay at the input, the rapid changing control signal will be
distorted a lot which will affect the feedback and then may
destabilize the system. This explains why high-gain feedback is
often avoided in practice.

Compared with these existing literature, the problem studied in this
paper is more general since not only the uncertainties at the input
but also the output feedback and mismatching are considered. For
output feedback, the state needs to be estimated which is difficult
mainly due to the uncertainties and disturbances in the state
equation. Even if parameters and disturbance can be estimated, it is
also difficult to compensate for mismatching uncertain parameters
and disturbance directly. To tackle these difficulties, two new
mechanisms are adopted in this paper. First, the input is redefined
to make it smooth and bounded to handle uncertainties at input. As a
consequence, the effect of unmodeled high-frequency gains and time
delay at the input is always bounded. And then, to handle estimate
and mismatching problem,\ the input-redefinition system is
transformed to an uncertainty-free system, which is proved to be
input-output equivalent with the aid of the \emph{additive state
decomposition}\footnote{In this paper we have replaced the term
\textquotedblleft additive decomposition\textquotedblright\ in
\cite{Quan(2009)} with the more descriptive term \textquotedblleft
additive state decomposition\textquotedblright.}\emph{
}\cite{Quan(2009)}. All mismatching\textbf{ }uncertainties,
mismatching disturbance and effect of unmodeled dynamics plus time
delay are lumped into a new disturbance at the output. An observer
is then designed for the transformed system to estimate the new
state and the new disturbance. Next, the transformed system is
`additively' decomposed into two independent subsystems in charge of
corresponding subtasks, namely the tracking (including rejection)
subtask and the input-realization subtask. Then one can design
controller for each subtask respectively, and finally combines them
to achieve the original control task. Three benchmark examples are
given to demonstrate the effectiveness of the proposed control
scheme.

The additive state decomposition is a decomposition scheme also
proposed in our previous work \cite{Quan(submit)}, where the
additive state decomposition is used to transform output feedback
tracking control for systems with measurable nonlinearities and
unknown disturbances and then to decompose it into three simpler
problems. This hence makes a challenging control problem tractable.
In this paper, a different control problem is investigated by using
additive state decomposition. Correspondingly, the transform and
decomposition are different. The major contributions of this paper
are: i) a tracking control scheme proposed to handle mismatching
parametric uncertainty, mismatching additive disturbances, unmodeled
high-frequency gains and time delay at the input; ii) a model
transform proposed to lump various uncertainties together; iii)
additive state decomposition in the controller design, especially in
how to handle saturation term.

This paper is organized as follows. In Section II, the problem
formulation is given and the additive state decomposition is
introduced briefly first. In Section III, input is redefined and the
input-redefinition system is transformed to an uncertainty-free
system in sense of input-output equivalence. Sequently, controller
design is given in Section IV. In Section V, two-cart example is
revisited by the proposed control scheme. Section VI concludes this
paper.

\section{Problem Formulation and Additive State Decomposition}

\subsection{Problem Formulation}

Consider a class of SISO nonlinear systems as follows:%
\begin{align}
\dot{x}  &  =f\left(  t,x,\theta\right)  +bu_{\xi}+d,x\left(
0\right)
=x_{0}\nonumber\\
y  &  =c^{T}x. \label{perturbedsystem}%
\end{align}
Here $b\in%
\mathbb{R}
^{n}\ $and $c\in%
\mathbb{R}
^{n}\ $are constant vectors, $\theta\left(  t\right)  \in%
\mathbb{R}
^{m}$ belongs to a given compact set $\Omega\subseteq%
\mathbb{R}
^{m},$ $x\left(  t\right)  \in%
\mathbb{R}
^{n}$ is the state vector, $y\left(  t\right)  \in%
\mathbb{R}
\ $is the output,$\ d\left(  t\right)  \in%
\mathbb{R}
^{n}$ is a bounded disturbance vector, and $u_{\xi}\left(  t\right)  \in%
\mathbb{R}
$ is the control subject to an unmodeled high-frequency gain and a
time delay
as follows:%
\begin{equation}
u_{\xi}\left(  s\right)  =H\left(  s\right)  e^{-\tau s}u\left(
s\right)
\label{high-frequency gain}%
\end{equation}
where $H\left(  s\right)  $ is an unknown stable proper transfer
function with $H\left(  0\right)  =1$ representing the unmodeled
high-frequency gain at the
input and $\tau\in%
\mathbb{R}
$ is the input delay. It is assumed that only $y$ is available from
measurement. The desired trajectory $r\left(  t\right)  \in%
\mathbb{R}
$ is known a priori, $t\geq0$. In the following, for convenience, we
will drop the notation $t$ except when necessary for clarity.

For system (\ref{perturbedsystem}), the following assumptions are
made.

\textbf{Assumption 1}. The function $f$ $:\left[  0,\infty\right)  \times%
\mathbb{R}
^{n}\times%
\mathbb{R}
^{m}\rightarrow%
\mathbb{R}
^{n}$ satisfies $f\left(  t,0,\theta\right)  \equiv0,$ and is
bounded when $x$ is bounded on $\left[  0,\infty\right)  $.
Moreover, for given $\theta
\in\Omega,$ there exist positive definite matrices $P\in%
\mathbb{R}
^{n\times n}$ and $Q\in%
\mathbb{R}
^{n\times n}$ such that
\begin{equation}
P\partial_{x}f\left(  t,x,\theta\right)  +\partial_{x}^{T}f\left(
t,x,\theta\right)  P\leq-Q,\forall x\in%
\mathbb{R}
^{n}, \label{Assumption1}%
\end{equation}
where $\partial_{x}f\triangleq\frac{\partial f}{\partial x}\in%
\mathbb{R}
^{n\times n}.$

\textbf{Definition 1 }\cite{Cao(2008)}. The $\mathcal{L}_{1}$ gain
of a stable proper SISO system is defined $\left\Vert G\left(
s\right)  \right\Vert
_{\mathcal{L}_{1}}=%
{\displaystyle\int\nolimits_{0}^{\infty}}
\left\vert g\left(  t\right)  \right\vert $d$t,$ where $g\left(
t\right)  $ is the impulse response of $G\left(  s\right)  $.

\textbf{Assumption 2}\textit{. }There exists a known stable proper
transfer function $C\left(  s\right)  $ with $C\left(  0\right)  =1$
such that $\left\Vert C\left(  s\right)  \left(  H\left(  s\right)
-1\right) \right\Vert _{\mathcal{L}_{1}}$ $\leq\varepsilon_{H},$
$\left\Vert sC\left( s\right)  \right\Vert _{\mathcal{L}_{1}}$
$\leq\varepsilon_{\tau}$, where
$\varepsilon_{H},\varepsilon_{\tau}\in%
\mathbb{R}
$ are positive real.

Under \textit{Assumptions 1-2},\textit{\ }the objective here is to
design a tracking controller $u$ such that $y\rightarrow r$ with a
good tracking accuracy, i.e., $y-r$ is ultimately bounded by a small
value.

\textbf{Remark 1}. From \textit{Assumption 1},\textbf{ }since
$f\left( t,x,\theta\right)  =\partial_{x}f\left(  t,x+\mu
x,\theta\right)  x,\mu \in\left(  0,1\right)  $ by the Taylor
expansion. Consequently, the system $\dot{x}=f\left(
t,x,\theta\right)  $ is exponentially stable by (\ref{Assumption1}).
In practice, many systems are stable themselves or they can be
stabilized by output feedback control. The following three benchmark
systems all satisfy \textit{Assumption 1}.

\textbf{Example 1 (}\textit{Rohrs' Example}\textbf{)}. Consider the
Rohrs'
example system as follows \cite{Rohrs(1982)}:%
\begin{equation}
y\left(  s\right)  =\frac{2}{s+1}\frac{229}{s^{2}+30s+229}u\left(
s\right)  .
\label{Rohrs0}%
\end{equation}
The nominal system is assumed to be $y\left(  s\right)
=\frac{2}{s+3}u\left( s\right)  $ here. In this case, the system
(\ref{Rohrs0}) can be formulated
into (\ref{perturbedsystem}) as%
\begin{align}
\dot{x}  &  =-\left(  3+\theta\right)  x+2u_{\xi},x\left(  0\right)
=1\nonumber\\
y  &  =x \label{Rohrs}%
\end{align}
where the parameter $\theta=-2$ is assumed unknown and $H\left(
s\right) =\frac{229}{s^{2}+30s+229},\tau=0$. It is easy to see that
\textit{Assumption 1 }is satisfied. Choose $C\left(  s\right)
=\frac{1}{2s+1}.$ Then \textit{Assumption 2 }is satisfied with
$\varepsilon_{H}=0.12\ $and $\varepsilon_{\tau}=1.$

\textbf{Example 2 (}\textit{Nonlinear}\textbf{)}. Consider a simple
nonlinear
system as follows \cite{Hagenmeyer(2003)}:%
\begin{align}
\dot{x}  &  =-x-\left(  1+\theta\right)  x^{3}+u\left(
t-\tau\right)
+d,x\left(  0\right)  =1\nonumber\\
y  &  =x, \label{Nonlinear}%
\end{align}
where $x,y,u,d\in%
\mathbb{R}
,\ $the parameter $\theta\left(  t\right)  =0.2\sin\left(
0.1t+1\right)  $, the input delay $\tau=0.1$ and $d\left(  t\right)
=0.5\sin\left( 0.2t\right)  $ are assumed unknown. The system
(\ref{Nonlinear}) can be formulated into (\ref{perturbedsystem})
with $f\left(  t,x,\theta\right) =-x-\left(  1+\theta\right)  x^{3}$
and $H\left(  s\right)  =1,\tau=0.1.$ It is easy to verify
$\partial_{x}f\left(  t,x,\theta\right)  =-1-3\left( 1+\theta\right)
x^{2}\leq-1.$ Therefore, \textit{Assumption 1 }is satisfied. Let
$C\left(  s\right)  =\frac{1}{2s+1}.$ Then \textit{Assumption 2 }is
satisfied with $\varepsilon_{H}=0\ $and $\varepsilon_{\tau}=1.$

\textbf{Remark 2}. The Rohrs' example system in \textit{Example 1
}is proposed to demonstrate that conventional adaptive control
algorithms developed at that time lose their robustness in the
presence of unmodeled dynamics \cite{Rohrs(1982)}. For the tracking
problem in \textit{Example 2}, there exist robustness issues by
using exact feedback linearization \cite{Hagenmeyer(2003)}. Compared
with the system in \cite{Hagenmeyer(2003)}, the input delay is added
in (\ref{Nonlinear}) to make system worse. The two benchmark
examples tell us that the uncertainties either on the system
parameters or at the input cannot be ignored in practice when design
a tracking controller, even if the original systems are stable. This
is also the initial motivation of this paper.

\subsection{Additive State Decomposition}

In order to make the paper self-contained, additive state
decomposition \cite{Quan(2009)} is introduced briefly here. Consider
the following
`original' system:%
\begin{equation}
f\left(  {t,\dot{x},x}\right)  =0,x\left(  0\right)  =x_{0}
\label{Gen_Orig_Sys}%
\end{equation}
where $x\in%
\mathbb{R}
^{n}$. We first bring in a `primary' system having the same
dimension as
(\ref{Gen_Orig_Sys}), according to:%
\begin{equation}
f_{p}\left(  {t,\dot{x}_{p},x_{p}}\right)  =0,x{_{p}}\left(
0\right)
=x_{p,0} \label{Gen_Pri_Sys}%
\end{equation}
where ${x_{p}}\in%
\mathbb{R}
^{n}$. From the original system (\ref{Gen_Orig_Sys}) and the primary
system
(\ref{Gen_Pri_Sys}) we derive the following `secondary' system:%
\begin{equation}
f\left(  {t,\dot{x},x}\right)  -f_{p}\left(
{t,\dot{x}_{p},x_{p}}\right)
=0,x\left(  0\right)  =x_{0} \label{Gen_Sec_Sys0}%
\end{equation}
where ${x_{p}}\in%
\mathbb{R}
^{n}$ is given by the primary system (\ref{Gen_Pri_Sys}). Define a
new
variable ${x_{s}}\in%
\mathbb{R}
^{n}$ as follows:%
\begin{equation}
{x_{s}\triangleq x-x_{p}}. \label{Gen_RelationPS}%
\end{equation}
Then the secondary system (\ref{Gen_Sec_Sys0}) can be further
written as
follows:%
\begin{equation}
f\left(  {t,\dot{x}_{s}+\dot{x}_{p},x_{s}+x_{p}}\right)
-f_{p}\left( {t,\dot{x}_{p},x_{p}}\right)  =0,x{_{s}}\left(
0\right)  =x_{0}-x_{p,0}.
\label{Gen_Sec_Sys}%
\end{equation}
From the definition (\ref{Gen_RelationPS}), we have%
\begin{equation}
{x}\left(  t\right)  ={x_{p}\left(  t\right)  +x_{s}\left(  t\right)
,t\geq0.} \label{Gen_RelationPS1}%
\end{equation}

\textbf{Remark 3}\textit{.} By the additive state
decomposition\textit{, }the\textit{ }system (\ref{Gen_Orig_Sys}) is
decomposed into two subsystems with the same dimension as the
original system. In this sense our decomposition is
\textquotedblleft additive\textquotedblright. In addition, this
decomposition is with respect to state. So, we call it
\textquotedblleft additive state
decomposition\textquotedblright\emph{.}

As a special case of (\ref{Gen_Orig_Sys}), a class of differential
dynamic
systems is considered as follows:%
\begin{align}
\dot{x}  &  =f\left(  {t,x}\right)  ,x\left(  0\right)  =x_{0},\nonumber\\
y  &  =h\left(  {t,x}\right)  \label{Dif_Orig_Sys}%
\end{align}
where ${x}\in%
\mathbb{R}
^{n}$ and $y\in%
\mathbb{R}
^{m}.$ Two systems, denoted by the primary system and (derived)
secondary
system respectively, are defined as follows:%
\begin{align}
\dot{x}_{p}  &  =f_{p}\left(  {t,x_{p}}\right)  ,x_{p}\left(
0\right)
=x_{p,0}\nonumber\\
y_{p}  &  =h_{p}\left(  {t,x}_{p}\right)  \label{Dif_Pri_Sys}%
\end{align}
and%
\begin{align}
\dot{x}_{s}  &  =f\left(  {t,x_{p}}+{x_{s}}\right)  -f_{p}\left(  {t,x_{p}%
}\right)  ,x_{s}\left(  0\right)  =x_{0}-x_{p,0},\nonumber\\
y_{s}  &  =h\left(  {t,x_{p}}+{x_{s}}\right)  -h_{p}\left(
{t,x}_{p}\right)
\label{Dif_Sec_Sys}%
\end{align}
where ${x_{s}}\triangleq{x-x_{p}}$ and $y_{s}\triangleq{y-y_{p}}$.
The secondary system (\ref{Dif_Sec_Sys}) is determined by the
original system (\ref{Dif_Orig_Sys}) and the primary system
(\ref{Dif_Pri_Sys}). From the
definition, we have%
\begin{equation}
{x}\left(  t\right)  ={x_{p}\left(  t\right)  +x_{s}\left(  t\right)
,y\left(  t\right)  =y_{p}\left(  t\right)  +y_{s}\left(  t\right)
,t\geq0.}
\label{Gen_RelationDif}%
\end{equation}

\section{Input Redefinition and Model Transformation}

Since $H\left(  s\right)  $ is the unmodeled high-frequency gain and
$\tau$ is the input delay, the control signal should be smooth
(low-frequency signal) so that it will maintain its original form as
far as possible after passing $H\left(  s\right)  e^{-\tau s}$.
Otherwise, the control signal will be distorted a lot. This explains
why high-gain feedback in practice is often avoided. For such a
purpose, the input is redefined to make control signal smooth and
bounded first. This makes the effect of $H\left(  s\right) e^{-\tau
s}$ under control, i.e., the effect will be predicted and bounded.

\subsection{Input Redefinition}

Redefine the input as follows:%
\[
u\left(  s\right)  =C\left(  s\right)  \left[  \sigma_{a}\left(
v\right) \left(  s\right)  \right]
\]
where $v\in%
\mathbb{R}
$ is the redefined control input and $\sigma_{a}:%
\mathbb{R}
\rightarrow\left[  -a,a\right]  $ is a saturation function defined
as $\sigma_{a}\left(  x\right)  \triangleq$sign$\left(  x\right)
\min\left(
\left\vert x\right\vert ,a\right)  $. Then $u_{\xi}$ is written as%
\begin{align}
u_{\xi}\left(  s\right)   &  =H\left(  s\right)  e^{-\tau s}C\left(
s\right)
\left[  \sigma_{a}\left(  v\right)  \left(  s\right)  \right] \nonumber\\
&  =C\left(  s\right)  \left[  \sigma_{a}\left(  v\right)  \left(
s\right)
\right]  +\xi\left(  s\right)  \label{controller2}%
\end{align}
where $\xi\left(  s\right)  =C\left(  s\right)  \left(  H\left(
s\right) e^{-\tau s}-1\right)  \left[  \sigma_{a}\left(  v\right)
\left(  s\right) \right]  \ $represents the effect of the unmodeled
high-frequency gain and the
time delay. The function $\xi\left(  s\right)  $ can be further written as%
\begin{equation}
\xi\left(  s\right)  =C\left(  s\right)  \left(  H\left(  s\right)
-1\right) e^{-\tau s}\left[  \sigma_{a}\left(  v\right)  \left(
s\right)  \right] +C\left(  s\right)  \left(  e^{-\tau s}-1\right)
\left[  \sigma_{a}\left(
v\right)  \left(  s\right)  \right]  . \label{highfreq}%
\end{equation}
From the definition of $\sigma_{a}$, we have $\sup_{-\infty<x<\infty
}\left\vert \sigma_{a}\left(  x\right)  \right\vert \leq a.$ In this paper $%
\mathcal{L}%
^{-1}$ denotes the inverse Laplace transform. By \textit{Assumption
2}, $\xi$
is bounded as follows:%
\begin{align}
\sup_{t\geq0}\left\vert \xi\left(  t\right)  \right\vert  &
\leq\left\Vert C\left(  s\right)  \left(  H\left(  s\right)
-1\right)  \right\Vert
_{\mathcal{L}_{1}}a+\left\Vert sC\left(  s\right)  \right\Vert _{\mathcal{L}%
_{1}}\sup_{t\geq0}\left\vert
\mathcal{L}%
^{-1}\left\{  \left(  e^{-\tau s}-1\right)  \left/  s\right.  \left[
\sigma_{a}\left(  v\right)  \left(  s\right)  \right]  \right\}
\right\vert
\nonumber\\
&  \leq\varepsilon_{H}a+\varepsilon_{\tau}\sup_{t\geq0}\left\vert
{\displaystyle\int\nolimits_{t}^{t-\tau}}
\sigma_{a}\left(  v\left(  \lambda\right)  \right)
d\lambda\right\vert
\nonumber\\
&  \leq\left(  \varepsilon_{H}+\tau\varepsilon_{\tau}\right)  a
\label{highfreqbound}%
\end{align}
where $\xi\left(  t\right)  =%
\mathcal{L}%
^{-1}\left(  \xi\left(  s\right)  \right)  .$The input redefinition
makes $\xi$ bounded not matter what the redefined control input $v$
is. Therefore, the redefined control input $v$ can be designed
freely. According to input
redefinition above, the controller (\ref{high-frequency gain}) is rewritten as%
\begin{equation}
u_{\xi}=u+\xi. \label{controller0}%
\end{equation}
Here $u\left(  t\right)  =%
\mathcal{L}%
^{-1}\left(  C\left(  s\right)  \left[  \sigma_{a}\left(  v\right)
\left( s\right)  \right]  \right)  $ can be written in the form of
state equation as
follows%
\begin{align}
\dot{z}  &  =A_{z}z+b_{z}\sigma_{a}\left(  v\right) \nonumber\\
u  &  =c_{z}^{T}z+d_{z}\sigma_{a}\left(  v\right)  \label{controller}%
\end{align}
where the vectors and matrices are compatibly dimensioned depending
on $C\left(  s\right)  .$ Substituting (\ref{controller0}) into the
system
(\ref{perturbedsystem}) results in%
\begin{align}
\dot{x}  &  =f\left(  t,x,\theta\right)  +bu+d_{h},x\left(  0\right)
=x_{0}\nonumber\\
y  &  =c^{T}x \label{perturbedsyste1}%
\end{align}
where $d_{h}=d+\xi.$ The system (\ref{perturbedsyste1}) with the
redefined controller (\ref{controller}) is called as the
\emph{input-redefinition system }here.

\subsection{Model Transformation}

The unknown parameter $\theta$ and the unknown disturbances $d$ are
not appear in \textquotedblleft matching\textquotedblright\
positions for the control input, i.e., $\theta$ and $d$ do not
appear like $b\left(  u_{\xi}+\theta ^{T}x+d\right)  .$ Therefore,
in a general system except for one dimensional system, the unknown
uncertainties cannot be often compensated for directly. Even if
$\theta$ and $d$ satisfy the \textquotedblleft matching
condition\textquotedblright, it is also difficult to compensate for
since the state $x$ is unknown. To tackle this difficulty, we first
transform the input-redefinition system (\ref{perturbedsyste1}) to
an uncertainty-free system, which is proved to be input-output
equivalent with the aid of the additive state decomposition as
stated in \textit{Theorem 1}. Before proving the theorem, the
following lemma is needed.

\textbf{Lemma 1}. Consider the following system%
\begin{equation}
\dot{x}=f\left(  t,x+z,\theta\right)  -f\left(  t,z,\theta\right)
+\rho\label{consys}%
\end{equation}
where $\rho\left(  t\right)  \in%
\mathbb{R}
^{n}$ is bounded. Under \textit{Assumption 1}, the solutions of
(\ref{consys})
satisfy%
\begin{equation}
\left\Vert x\left(  t\right)  \right\Vert \leq\beta\left(
\left\Vert x\left( t_{0}\right)  \right\Vert ,t-t_{0}\right)
+\gamma\underset{t_{0}\leq s\leq
t}{\sup}\left\Vert \rho\left(  s\right)  \right\Vert \label{consysBIBS}%
\end{equation}
where $\beta$ is a class $\mathcal{KL}$ function
\cite[p.144]{Khalil(2002)} and
$\gamma=2\frac{\lambda_{\max}^{2}\left(  P\right)
}{\lambda_{\min}\left( P\right)  \lambda_{\min}\left(  Q\right)  }$.

\textit{Proof}. By the Taylor expansion, the function $f\left(
t,x+z,\theta
\right)  $ can be written as%
\[
f\left(  t,x+z,\theta\right)  =f\left(  t,z,\theta\right)  +\partial
_{x}f\left(  t,x+z+\mu x,\theta\right)  x
\]
where $\mu\in\left(  0,1\right)  .$ Then the system (\ref{consys})
can be
rewritten as%
\begin{equation}
\dot{x}=\partial_{x}f\left(  t,x+z+\mu x,\theta\right)  x+\rho.
\label{consys1}%
\end{equation}
Choose Lyapunov function $V=x^{T}Px.$ By \textit{Assumption 1}, the
derivative
of $V$ along (\ref{consys1}) satisfies%
\begin{align*}
\dot{V}  &  \leq-\lambda_{\min}\left(  Q\right)  \left\Vert
x\right\Vert ^{2}+\lambda_{\max}\left(  P\right)  \left\Vert
x\right\Vert \left\Vert
\rho\right\Vert \\
&  \leq-\frac{1}{2}\lambda_{\min}\left(  Q\right)  \left\Vert
x\right\Vert ^{2},\text{ }\forall\left\Vert x\right\Vert
\geq2\frac{\lambda_{\max}\left( P\right)  }{\lambda_{\min}\left(
Q\right)  }\left\Vert \rho\right\Vert .
\end{align*}
By Theorem 4.19 \cite[p.176]{Khalil(2002)}, we can conclude this
proof. $\square$

With \textit{Lemma 1} in hand, we have

\textbf{Theorem 1}\textit{.} Under \textit{Assumption 1}, there
always exists an estimate of $\theta,$ namely
$\hat{\theta}\in\Omega,$ such that the system
(\ref{perturbedsyste1}) is input-output equivalent to the following system:%
\begin{align}
\dot{x}_{new}  &  =f(t,x_{new},\hat{\theta})+bu,x_{new}\left(
0\right)
=0\nonumber\\
y  &  =c^{T}x_{new}+d_{new}. \label{equ1_equivalent}%
\end{align}
Here $x_{new}$ and $d_{new}$ satisfy%
\begin{align}
\left\Vert x-x_{new}\right\Vert  &  \leq\beta\left(  \left\Vert x_{0}%
\right\Vert ,t-t_{0}\right)  +\gamma\underset{t_{0}\leq s\leq
t}{\sup
}\left\Vert d_{\tilde{\theta}}\left(  s\right)  \right\Vert \nonumber\\
\left\Vert d_{new}\right\Vert  &  \leq\left\Vert c\right\Vert
\beta\left( \left\Vert x_{0}\right\Vert ,t-t_{0}\right)  +\left\Vert
c\right\Vert \gamma\underset{t_{0}\leq s\leq t}{\sup}\left\Vert
d_{\tilde{\theta}}\left(
s\right)  \right\Vert \label{equ1_equivalent_pro}%
\end{align}
where $\beta$ is a class $\mathcal{KL}$ function,
$\gamma=2\frac{\lambda _{\max}^{2}\left(  P\right)
}{\lambda_{\min}\left(  P\right)  \lambda_{\min
}\left(  Q\right)  }$ and $d_{\tilde{\theta}}=f\left(  t,x_{new}%
,\theta\right)  -f(t,x_{new},\hat{\theta})+d_{h}.$

\textit{Proof}. In the following, additive state decomposition is
utilized to decompose the system (\ref{perturbedsyste1}) first.
Consider the system (\ref{perturbedsyste1}) as the original system
and choose the primary system
as follows:%
\begin{align}
\dot{x}_{p}  &  =f(t,x_{p},\hat{\theta})+bu,x_{p}\left(  0\right)
=0\nonumber\\
y_{p}  &  =c^{T}x_{p}. \label{equ1_Pri0}%
\end{align}
Then the secondary system is determined by the original system
(\ref{perturbedsyste1}) and the primary system (\ref{equ1_Pri0})
with the rule
(\ref{Dif_Sec_Sys}) that%
\begin{align}
\dot{x}_{s}  &  =f\left(  t,x_{p}+x_{s},\theta\right)
-f(t,x_{p},\hat{\theta
})+d_{h},x_{s}\left(  0\right)  =x_{0}\nonumber\\
y_{s}  &  =c^{T}x_{s}. \label{equ1_Sec0}%
\end{align}
According to (\ref{Gen_RelationDif}), we have $x=x_{p}+x_{s}\ $and$\ y=y_{p}%
+y_{s}.$ Consequently, we can get an uncertainty-free system as follows%
\begin{align*}
\dot{x}_{p}  &  =f(t,x_{p},\hat{\theta})+bu,x_{p}\left(  0\right)  =0\\
y  &  =c^{T}x_{p}+y_{s}%
\end{align*}
where $u$ and $y$ are the same to those in (\ref{perturbedsyste1}).
Let $x_{p}=x_{new}$ and $d_{new}=y_{s}.$ We can conclude that the
system (\ref{perturbedsyste1}) is input-output equivalent to
(\ref{equ1_equivalent}). Next, we will prove that
(\ref{equ1_equivalent_pro}) is satisfied. The system
(\ref{equ1_Sec0}) can be rewritten as%
\begin{align}
\dot{x}_{s}  &  =f\left(  t,x_{p}+x_{s},\theta\right)  -f\left(
t,x_{p},\theta\right)  +d_{\tilde{\theta}},x_{s}\left(  0\right)
=x_{0}\nonumber\\
y_{s}  &  =c^{T}x_{s} \label{equ1_Sec1}%
\end{align}
where $d_{\tilde{\theta}}=f\left(  t,x_{p},\theta\right)  -f(t,x_{p}%
,\hat{\theta})+d_{h}$. Then, by \textit{Lemma 1}, we have%
\begin{align*}
\left\Vert x\left(  t\right)  -x_{new}\left(  t\right)  \right\Vert
& =\left\Vert x_{s}\left(  t\right)  \right\Vert \leq\beta\left(
\left\Vert x_{0}\right\Vert ,t-t_{0}\right)
+\gamma\underset{t_{0}\leq s\leq t}{\sup
}\left\Vert d_{\tilde{\theta}}\left(  s\right)  \right\Vert \\
\left\Vert d_{new}\left(  t\right)  \right\Vert  &  \leq\left\Vert
c\right\Vert \left\Vert x_{s}\left(  t\right)  \right\Vert
\leq\left\Vert c\right\Vert \beta\left(  \left\Vert x_{0}\right\Vert
,t-t_{0}\right) +\left\Vert c\right\Vert \gamma\underset{t_{0}\leq
s\leq t}{\sup}\left\Vert d_{\tilde{\theta}}\left(  s\right)
\right\Vert .
\end{align*}
$\square$

For the uncertainty-free transformed system (\ref{equ1_equivalent}),
we design an observer to estimate $x_{new}$ and $d_{new}$, which is
stated in \textit{Theorem 2.}

\textbf{Theorem 2}\textit{.} Under \textit{Assumption 1}, an
observer is designed to estimate state $x_{new}$ and $d_{new}$ in
(\ref{equ1_equivalent})
as follows%
\begin{align}
\dot{\hat{x}}_{new}  &  =f(t,\hat{x}_{new},\hat{\theta})+bu,\hat{x}%
_{new}\left(  0\right)  =0\nonumber\\
\hat{d}_{new}  &  =y-c^{T}\hat{x}_{new}. \label{equ1_equivalent_est}%
\end{align}
Then $\hat{x}_{new}\equiv x_{new}$ and $\hat{d}_{new}\equiv
d_{new}.$

\textit{Proof.} Subtracting (\ref{equ1_equivalent_est}) from
(\ref{equ1_equivalent}) results in
\[
\dot{\tilde{x}}_{new}=\partial_{x}f(t,x_{new}+\hat{x}_{new}+\mu x_{new}%
,\hat{\theta})\tilde{x}_{new},\tilde{x}_{new}\left(  0\right)  =0,
\]
where $\mu\in\left(  0,1\right)  $ and$\ \tilde{x}_{new}\triangleq
x_{new}-\hat{x}_{new}.$ Then $\tilde{x}_{new}\equiv0$. This implies
that
$\hat{x}_{new}\equiv x_{new}.$ Consequently, by the relation $y=c^{T}%
x_{new}+d_{new}$ in (\ref{equ1_equivalent}), we have
$\hat{d}_{new}\equiv d_{new}.$ $\square$

\textbf{Remark 4}\textit{. }By\textit{ }(\ref{controller}), the
control signal $u$ is always bounded. Therefore, by \textit{Lemma
1}, the state\textit{
}$x_{new}$ is always bounded. Consequently, by\textit{\ }%
(\ref{equ1_equivalent_pro}), $d_{new}$ is always bounded as well. It
is
interesting to note that the new state\textit{\ }$x_{new}$ and\textit{\ }%
disturbance $d_{new}$\textit{\ }in the transformed system
(\ref{equ1_equivalent}) can be observed directly rather than
asymptotically or exponentially. This will facilitate the analysis
and design later.

\textbf{Example 3 (}\textit{Rohrs' Example, Example 1
Continued}\textbf{)}. According to input redefinition above, the
Rohrs' example system (\ref{Rohrs})
can be rewritten as follows:%
\begin{align*}
\dot{x}  &  =-\left(  3+\theta\right)  x+2u+\left(  d+2\xi\right) \\
y  &  =x
\end{align*}
where $\sup_{t\geq0}\left\vert \xi\left(  t\right)  \right\vert
\leq0.12a,$ and $u$ is generated by
$\dot{z}=-0.5z+0.5\sigma_{a}\left(  v\right)  ,u=z.$ Then, according
to (\ref{equ1_equivalent}), the uncertainty-free transformed
system of (\ref{Rohrs}) is%
\begin{align}
\dot{x}_{new}  &  =-(3+\hat{\theta})x_{new}+2u\nonumber\\
y  &  =x_{new}+d_{new} \label{Rohrs_Tran}%
\end{align}
where $\hat{\theta}\ $will be specified later.

\textbf{Example 4 (}\textit{Nonlinear, Example 2
Continued}\textbf{)}. According to input redefinition above, the
nonlinear system (\ref{Nonlinear})
can be rewritten as follows:%
\begin{align*}
\dot{x}  &  =-x-\left(  1+\theta\right)  x^{3}+u+\left(  d+\xi\right) \\
y  &  =x
\end{align*}
where $\sup_{t\geq0}\left\vert \xi\left(  t\right)  \right\vert
\leq0.1a,$ and $u$ is generated by
$\dot{z}=-0.5z+0.5\sigma_{a}\left(  v\right)  ,u=z.$ Then, according
to (\ref{equ1_equivalent}), the uncertainty-free transformed system
of (\ref{Nonlinear}) is%
\begin{align}
\dot{x}_{new}  &
=-x_{new}-(1+\hat{\theta})x_{new}^{3}+u,x_{new}\left(
0\right)  =0\nonumber\\
y  &  =x_{new}+d_{new} \label{Nonlinear_Tran}%
\end{align}
where $\hat{\theta}\ $will be specified later.

\section{Controller Design}

In this section, the transformed system (\ref{equ1_equivalent}) is
`additively' decomposed into two independent subsystems in charge of
corresponding subtasks. Then one can design controller for each
subtask respectively, and finally combines them to achieve the
original control task.

\subsection{Additive State Decomposition of Transformed System}

Currently, based on the new transformed system
(\ref{equ1_equivalent}), the objective is to design a tracking
controller $u$ such that $y\rightarrow r$ with a good tracking
accuracy, i.e., $y-r$ is ultimately bounded by a small value. While,
$u$ is realized by (\ref{controller}). According to this fact, the
transformed system (\ref{equ1_equivalent}) is `additively'
decomposed into two independent subsystems in charge of
corresponding subtasks, namely the tracking (including rejection)
subtask and the input-realization subtask. This is shown in
Fig.1.\begin{figure}[h]
\begin{center}
\includegraphics[
scale=0.9 ]{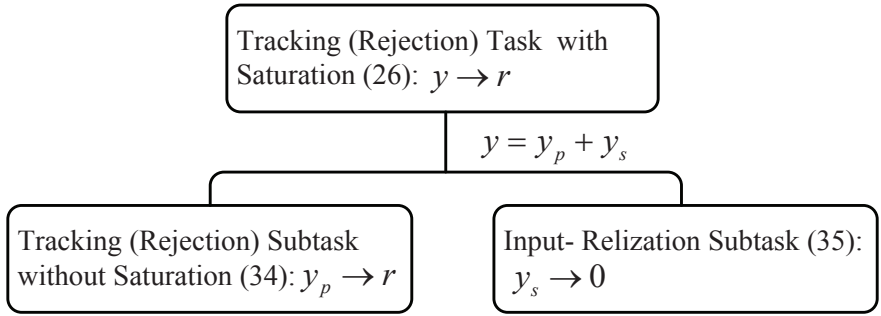}
\end{center}
\caption{Additive state decomposition flow}%
\end{figure}

Consider the transformed system (\ref{equ1_equivalent}) as the
original system. According to the principle above, we choose the
primary system as
follows:%
\begin{align}
\dot{x}_{p}  &  =f(t,x_{p},\hat{\theta})+bu_{p},x_{p}\left(
0\right)
=0\nonumber\\
y_{p}  &  =c^{T}x_{p}+d_{new}. \label{equ1_Pri}%
\end{align}
Then the secondary system is determined by the original system
(\ref{equ1_equivalent}) and the primary system (\ref{equ1_Pri}) with
the rule
(\ref{Dif_Sec_Sys}), and we can obtain that%
\begin{align}
\dot{x}_{s}  &  =f(t,x_{p}+x_{s},\hat{\theta})-f(t,x_{p},\hat{\theta
})+b\left(  u-u_{p}\right)  ,x_{s}\left(  0\right)  =0\nonumber\\
y_{s}  &  =c^{T}x_{s}. \label{equ1_Sec}%
\end{align}
According to (\ref{Gen_RelationDif}), we have%
\begin{equation}
x_{new}=x_{p}+x_{s}\text{ and }y=y_{p}+y_{s}. \label{equ1_relation}%
\end{equation}

The strategy here is to assign the tracking (including rejection)
subtask to the primary system (\ref{equ1_Pri}) and the
input-realization subtask to the
secondary system (\ref{equ1_Sec}). It is clear from (\ref{equ1_Pri}%
)-(\ref{equ1_relation}) that if the controller $u_{p}$ drives $y_{p}%
\rightarrow r$ in (\ref{equ1_Pri}) and $u$ drives
$y_{s}\rightarrow0$ in (\ref{equ1_Sec}), then $y\rightarrow r$ as
$t\rightarrow\infty$. The benefit brought by the additive state
decomposition is that the controller $u$ will not affect the
tracking and rejection performance since the primary system
(\ref{equ1_Pri}) is independent of the secondary system
(\ref{equ1_Sec}). Since the states $x_{p}$ and $x_{s}$ are unknown
except for addition of them, namely $x_{new}$, an observer is
proposed to estimate $x_{p}$ and $x_{s}.$

\textbf{Remark 5}\textit{. }Although the proposed additive state
decomposition gives clear how to decompose a system, it still leaves
a freedom to choose the primary system. By the additive state
decomposition, the transformed system
(\ref{equ1_equivalent}) can be also decomposed into a primary system%
\begin{align}
\dot{x}_{p} &  =Ax_{p}+bu_{p},x_{p}\left(  0\right)  =0\nonumber\\
y_{p} &  =c^{T}x_{p}+d_{new}\label{equ1_Pri_1}%
\end{align}
and the derived secondary system%
\begin{align}
\dot{x}_{s} &  =f\left(  t,x_{p}+x_{s},\hat{\theta}\right)
-Ax_{p}+b\left(
u_{\xi}-u_{p}\right)  ,x_{s}\left(  0\right)  =x_{0}\nonumber\\
y_{s} &  =c^{T}x_{s}\label{equ1_Sec_1}%
\end{align}
where $A\in%
\mathbb{R}
^{n\times n}$ is an arbitrary constant matrix. Therefore, there is
an infinite number of decompositions. The principle here is to
derive the secondary system with an equilibrium point close to zero
as far as possible. If so, the problem for the secondary system is
only a stabilization problem, which is easier compared with a
tracking problem. In (\ref{equ1_Sec}), $x_{s}=0$ is an
equilibrium point of $\dot{x}_{s}=f(t,x_{p}+x_{s},\hat{\theta})-f(t,x_{p}%
,\hat{\theta}),$ whereas in (\ref{equ1_Sec_1}),$\ x_{s}=0$ is not an
equilibrium point of $\dot{x}_{s}=f\left(
t,x_{p}+x_{s},\theta\right) -Ax_{p}.$ This is why we choose the
primary system as (\ref{equ1_Pri}) not (\ref{equ1_Pri_1}). From the
mention above, a good additive state decomposition often depends on
a concrete problem.

\textbf{Theorem 3}\textit{. }Under \textit{Assumption 1}, suppose
that an
observer is designed to estimate state $x_{p}$ and $x_{s}$ in (\ref{equ1_Pri}%
)-(\ref{equ1_Sec}) as follows:
\begin{subequations}
\label{equ1_Obs}%
\begin{align*}
\dot{\hat{x}}_{p}  &
=f(t,\hat{x}_{p},\hat{\theta})+bu_{p},\hat{x}_{p}\left(
0\right)  =0\\
\hat{x}_{s}  &  =x_{new}-\hat{x}_{p}.
\end{align*}
Then $\hat{x}_{p}\equiv x_{p}$ and $\hat{x}_{s}\equiv x_{s}.$

\textit{Proof. }Similar to the proof of \textit{Theorem 2.}
$\square$

So far, we have transformed the original system to an
uncertainty-free system, in which the new state and the new
disturbance can be estimated directly. And then, decompose the
transformed system into two independent subsystems in charge of
corresponding subtasks. In the following, we are going to
investigate the controller design with respect to the two decomposed
subtasks respectively.

\subsection{Problem for Tracking (including Rejection) Subtask}

\textbf{Problem 1}. \textit{For (\ref{equ1_Pri}), design a controller}%
\end{subequations}
\begin{equation}
u_{p}=u^{r}\left(  t,x_{p},r,d_{new}\right)  \label{Assumption2_Con}%
\end{equation}
\textit{such that }$y_{p}\rightarrow r+\mathcal{B}\left(
\delta_{r}\right)  $
\footnote{$\mathcal{B}\left(  \delta\right)  \triangleq\left\{  x\in%
\mathbb{R}
\left\vert \text{ }\left\vert x\right\vert \leq\delta\right.
\right\}  ;$ the notation $x\left(  t\right)
\rightarrow\mathcal{B}\left(  \delta\right)  $ means
$\underset{y\in\mathcal{B}\left(  \delta\right)  }{\min}$
$\left\Vert x\left(  t\right)  -y\right\Vert \rightarrow0;$
$\mathcal{B}\left(  \delta _{1}\right)  +\mathcal{B}\left(
\delta_{2}\right)  \triangleq\left\{  \left.
x+y\right\vert x\in\mathcal{B}\left(  \delta_{1}\right)  ,y\in\mathcal{B}%
\left(  \delta_{2}\right)  \right\}  $}\textit{\ as
}$t\rightarrow\infty ,$\textit{\ meanwhile keeping the state
}$x_{p}$\textit{ bounded, where
}$\delta_{r}\in%
\mathbb{R}
.$

\textbf{Remark 6}\textit{\ (on Problem 1)}. Since
$y_{p}=c^{T}x_{p}+d_{new},$ \textit{Problem 1} can be also
considered to design $u_{p}$\textit{ }such that
$c^{T}x_{p}\rightarrow r-d_{new}.$ Here, the difference between $r$
and $d_{new}$ should be clarified. The reference $r$ is often known
a priori, i.e., $r\left(  t+T\right)  $ is known at the time $t,$
where $T>0.$ Moreover, its derivative is often given or can be
obtained by analytic methods. Whereas, the new disturbance $d_{new}$
only can be obtained at the time $t\ $whose derivative only can be
obtained by numerical methods. By recalling
(\ref{equ1_equivalent_pro}), the new disturbance $d_{new}$ depends
on the disturbance $d,$ the parameters $\theta$ and $\hat{\theta},$
the effect of unmodeled high-frequency gain namely $\xi,$ the state
$x_{new},$ and initial value $x_{0}.$ One way of reducing the
complexity is to design an observer to estimate $\theta,\ $and makes
$\hat{\theta}\rightarrow\theta$ as $t\rightarrow\infty.$ As a
result, the new disturbance $d_{new}$ finally depends on $d$ and
$\xi\ $as $t\rightarrow\infty.$ In practice, low frequency band is
often dominant in the reference signal and disturbance. Therefore,
from a practical point of view,\ we can also modify the tracking
target, namely $r-d_{new}.$ For example, let $r-d_{new}$ pass a
low-pass filter to obtain its major component. If the major
component of $r-d_{new}$ belongs to a fixed family of functions of
time, \textit{Problem 1 }can also be considered as an output
regulation problem \cite{Isdori (2003)}.

\subsection{Problem for Input-Realization Subtask}

As shown in Fig.1, the input realization subtask aims to make $y_{s}%
\rightarrow0.$ Let us investigate the secondary system
(\ref{equ1_Sec}). By
\textit{Lemma 1}, we have%
\begin{equation}
\left\Vert x_{s}\left(  t\right)  \right\Vert \leq\beta\left(
\left\Vert x_{s}\left(  t_{0}\right)  \right\Vert ,t-t_{0}\right)
+\gamma\left\Vert b\right\Vert \underset{t_{0}\leq s\leq
t}{\sup}\left\Vert u\left(  s\right)
-u_{p}\left(  s\right)  \right\Vert . \label{equ1_Sec_rela}%
\end{equation}
This implies that $y_{s}\rightarrow\gamma\left\Vert b\right\Vert
\left\Vert c\right\Vert \mathcal{B}\left(  \delta_{s}\right)  $ as
$u\rightarrow
u_{p}+\mathcal{B}\left(  \delta_{s}\right)  $, where $\delta_{s}\in%
\mathbb{R}
.$ It is noticed that $u$ only can be realized by
(\ref{controller})\textit{. }Therefore, problem for
input-realization subtask can be stated as follows:

\textbf{Problem 2}. \textit{Given a signal }$u_{p},$\textit{ design
a controller }$v=v^{s}\left(  t,u_{p}\right)  $\textit{ for
(\ref{controller}) such that }$u\rightarrow u_{p}+\mathcal{B}\left(
\delta_{s}\right) $\textit{\ as }$t\rightarrow\infty.$

This is also a tracking problem but with a saturation constraint.
Here we give a solution to the \textit{Problem 2}. The main
difficult is how to handle the saturation in (\ref{controller}).
Here, additive state decomposition will be used again. Taking
(\ref{controller}) as the original system, we choose the
primary system as follows%
\begin{align}
\dot{z}_{p}  &  =A_{z}z_{p}+b_{z}v\nonumber\\
u_{zp}  &  =c_{z}^{T}z_{p}+d_{z}v \label{controller_Pri}%
\end{align}
Then the secondary system is determined by the original system
(\ref{controller}) and the primary system (\ref{controller_Pri})
with the rule
(\ref{Dif_Sec_Sys}), and we can obtain that%
\begin{align}
\dot{z}_{s}  &  =A_{z}z_{s}+b_{z}\left(  \sigma_{a}\left(  v\right)
-v\right)
\nonumber\\
u_{zs}  &  =c_{z}^{T}z_{p}+d_{z}\left(  \sigma_{a}\left(  v\right)
-v\right)
. \label{controller_Sec}%
\end{align}
According to (\ref{Gen_RelationDif}), we have $z=z_{p}+z_{s}$ and
$u=u_{zp}+u_{zs}.$ The benefit brought by the additive state
decomposition is that the controller saturation will not affect the
primary system (\ref{controller_Pri}). Moreover, the controller $v$
can be designed only based on the primary system
(\ref{controller_Pri}), where the controller $v$ uses the state
$z_{p}$ not $z.$ So, the strategy here is to design $v=v^{s}\left(
t,u_{p}\right)  $ in (\ref{controller_Pri}) to drive
$u_{zp}\rightarrow u_{p}$ as $t\rightarrow\infty$ and neglect the
secondary system (\ref{controller_Sec}). Since $v^{s}\left(
t,u_{p}\right)  $ is bounded, the state of the secondary system
(\ref{controller_Sec}) will be bounded as well. If $\sigma_{a}\left(
v^{s}\left(  t,u_{p}\right)  \right) -v^{s}\left(  t,u_{p}\right)
\rightarrow0$ as $t\rightarrow\infty,$ then $u_{zs}\rightarrow0$ as
$t\rightarrow\infty.$ Consequently, $u\rightarrow u_{p}$ as
$t\rightarrow\infty.$ For (\ref{controller_Pri}), the transfer
function from $v$ to $u_{zp}$ is $u_{zp}\left(  s\right)  =C\left(
s\right) v\left(  s\right)  .$ If $C\left(  s\right)  $ is designed
to be minimum phase, an easy way is to design $v$ to be
\begin{equation}
v\left(  s\right)  =C^{-1}\left(  s\right)  u_{p}\left(  s\right)  .
\label{easycontrol}%
\end{equation}
The design will make the signal $\sigma_{a}\left(  v\right)  $ close
to the idea one, meanwhile maintaining the signal $\sigma_{a}\left(
v\right)  $ smooth as far as possible. By recalling
(\ref{highfreq}), it will make the effect of the unmodeled
high-frequency gain and the time delay $\xi$ smaller.

\subsection{Controller Integration}

With the solutions of the two problems in hand, we can state

\textbf{Theorem 4}. Under \textit{Assumptions 1-2},\textit{\
}suppose i) \textit{Problems 1-2} are solved; ii) the controller for
system (\ref{perturbedsystem}) (or (\ref{equ1_equivalent})) is
designed as

Observer:%
\begin{align}
\dot{\hat{x}}_{new}  &  =f(t,\hat{x}_{new},\hat{\theta})+bu,\hat{x}%
_{new}\left(  0\right)  =0,\nonumber\\
\dot{\hat{x}}_{p}  &
=f(t,\hat{x}_{p},\hat{\theta})+bu_{p},\hat{x}_{p}\left(
0\right)  =0,\nonumber\\
\hat{d}_{new}  &  =y-c^{T}\hat{x}_{new} \label{mainobsever1}%
\end{align}

Controller:%
\begin{align}
u_{p}  &  =u^{r}(t,\hat{x}_{p},r,\hat{d}_{new}),v=v^{s}\left(
t,u_{p}\right)
\nonumber\\
\dot{z}  &  =A_{z}z+b_{z}\sigma_{a}\left(  v\right)  ,u=c_{z}^{T}z+d_{z}%
\sigma_{a}\left(  v\right)  \label{maincontroller1}%
\end{align}
Then the output of system (\ref{perturbedsystem}) (or
(\ref{equ1_equivalent})) satisfies that $y\rightarrow
r+\mathcal{B}\left(  \delta_{r}+\gamma\left\Vert b\right\Vert
\left\Vert c\right\Vert \delta_{s}\right)  $ as $t\rightarrow
\infty,$ meanwhile keeping all states bounded. In particular, if
$\delta _{r}+\delta_{s}=0,$ then the output in system
(\ref{perturbedsystem}) (or (\ref{equ1_equivalent})) satisfies that
$y\rightarrow r$ as $t\rightarrow \infty.$

\textit{Proof. }It is easy to follow the proof in \textit{Theorems
2-3 }that the\textit{\ }observer (\ref{mainobsever1}) will make
\begin{equation}
\hat{x}_{new}\equiv x_{new},\hat{d}_{new}\equiv
d_{new},\hat{x}_{p}\equiv
x_{p}. \label{mainobseverresult}%
\end{equation}
Suppose that \textit{Problem 1} is solved. By\
(\ref{Assumption2_Con}) and\textit{\ }(\ref{mainobseverresult}), the
controller $u_{p}=u^{r}(t,\hat {x}_{p},r,\hat{d}_{new})$ can drive
$y_{p}\rightarrow r+\mathcal{B}\left( \delta_{r}\right)  $ as
$t\rightarrow\infty$ in (\ref{equ1_Pri}). Suppose that
\textit{Problem 2} is solved. By\textit{ }(\ref{mainobseverresult}),
the controller $v=v^{s}\left(  t,u_{p}\right)  $ can drive
$u\rightarrow u_{p}+\mathcal{B}\left(  \delta_{s}\right)  $ as
$t\rightarrow\infty$ in
(\ref{equ1_Sec}). Further by (\ref{equ1_Sec_rela}), we have $y_{s}%
\rightarrow\mathcal{B}\left(  \gamma\left\Vert b\right\Vert
\left\Vert c\right\Vert \delta_{s}\right)  .$ Since $y=y_{p}+y_{s},$
we have $y\rightarrow r+\mathcal{B}\left(
\delta_{r}+\gamma\left\Vert b\right\Vert \left\Vert c\right\Vert
\delta_{s}\right)  .$ $\square$

\textbf{Example 5 (}\textit{Rohrs' Example, Example 3
Continued}\textbf{)}. According to (\ref{equ1_Pri}), the primary
system of linear system
(\ref{Rohrs_Tran}) can be rewritten as follows:%
\begin{align*}
\dot{x}_{p}  &  =-(3+\hat{\theta})x_{p}+2u_{p}\\
y  &  =x_{p}+d_{new}%
\end{align*}
Design
$u_{p}=\frac{1}{2}[(2+\hat{\theta})x_{p}+r+\dot{r}-d_{new}-\dot
{d}_{new}].$ Then the system above becomes $\dot{e}_{p}=-e_{p},$
where $e_{p}=y_{p}-r.$ Therefore, $y_{p}\rightarrow r$ as
$t\rightarrow\infty.$ According to (\ref{easycontrol}), $v$ is
designed as $v^{s}\left( t,u_{p}\right)  =2\dot{u}_{p}+u_{p}.$ Here
$\dot{u}_{p}$ and $\dot{d}_{new}$
are approximated by $\dot{d}_{new}\approx%
\mathcal{L}%
^{-1}(\frac{s}{0.1s+1}d_{new}\left(  s\right)  )$ and $\dot{u}_{p}\approx%
\mathcal{L}%
^{-1}(\frac{s}{0.1s+1}u_{p}\left(  s\right)  ),$ respectively.
Suppose $\hat{\theta}=0\ $and given $r=0.5$ and $r=0.5$sin$\left(
0.2t\right)  ,$ respectively. Driven by the resulting controller
(\ref{maincontroller1}), the simulation result is shown in Fig.2.

\textbf{Example 6 (}\textit{Nonlinear, Example 4
Continued}\textbf{)}. According to (\ref{equ1_Pri}), the primary
system of nonlinear system
(\ref{Nonlinear_Tran}) can be rewritten as follows:%
\begin{align*}
\dot{x}_{p}  &  =-x_{p}-(1+\hat{\theta})x_{p}^{3}+u_{p},x_{p}\left(
0\right)
=0\\
y_{p}  &  =x_{p}+d_{new}.
\end{align*}
Design
$u_{p}=(1+\hat{\theta})x_{p}^{3}+\dot{r}+r-\dot{d}_{new}-d_{new}.$
Then the system above becomes $\dot{e}_{p}=-e_{p}$, where
$e_{p}=y_{p}-r.$ Therefore, $y_{p}\rightarrow r$ as
$t\rightarrow\infty.$ According to (\ref{easycontrol}), $v^{s}\left(
t,u_{p}\right)  $ is designed as $v^{s}\left(  t,u_{p}\right)
=2\dot{u}_{p}+u_{p}.$ Here the derivative of
$u_{p}$ and $d_{new}$ are approximated by $\dot{d}_{new}\approx%
\mathcal{L}%
^{-1}(\frac{s}{0.1s+1}d_{new}\left(  s\right)  )$ and $\dot{u}_{p}\approx%
\mathcal{L}%
^{-1}(\frac{s}{0.1s+1}u_{p}\left(  s\right)  ),$ respectively.
Suppose $\hat{\theta}=0\ $and given $r=0.5$ and $r=0.5$sin$\left(
0.2t\right)  ,$ respectively. Driven by the resulting controller
(\ref{maincontroller1}), the simulation result is shown in Fig.3.

\textbf{Remark 7}. The derivative of $d_{new}$ and $u_{p}$ can be
also obtained by differentiator technique
\cite{Han(1994)},\cite{Levent(1998)}. It is interesting to note that
$\hat{\theta}$ is different from $\theta,$ but $y\rightarrow r$ with
a good tracking accuracy. This is one major advantage of this
proposed control scheme. Moreover, all the unknown parts such as
$\theta,d$ and the effect of $H\left(  s\right)  e^{-s\tau}$ are
treated as a lumped disturbance $d_{new}.$ This can explain why the
proposed scheme can handle many uncertainties.

\section{Two-Cart Example}

The two-cart mass-spring-damper example was originally proposed as a
benchmark problem for robust control design
\cite{Enric(2009)},\cite{Fekri(2006)}. Next, we will revisit the
two-cart example by the proposed control scheme.

The two-cart system is shown in Fig.4. The states $x_{1}(t)$ and
$x_{2}(t)$ represent the absolute position of the two carts, whose
masses are $m_{1}$ and
$m_{2}$ respectively; $k_{1},k_{2}$ are the spring constants, and $b_{1}%
,b_{2}$ are the damping coefficients; $d(t)$ is a disturbance force
acting on the mass $m_{2}$; $u(t)$ is the control force subject to
an unmodeled high-frequency gain and a time delay, which acts upon
the mass $m_{1}$. The parameter $m_{1}=1$ is known, whereas the
following parameters $m_{2}=2,$ $k_{1}=0.8,k_{2}=0.5,$ $b_{1}=1.3,$
$b_{2}=0.9$ are assumed unknown. The unmodeled high-frequency gain
and a time delay is assumed to be $H\left( s\right)  e^{-\tau
s}=\frac{229}{s^{2}+30s+229}e^{-0.1s}$. The disturbance force
$\zeta(t)$ is modeled as a first-order (colored) stochastic process
generated by driving a low-pass filter with continuous-time white
noise $\varepsilon(s)$, with zero-mean and unit intensity, i.e.
$\Xi=1$, as follows $\zeta\left(  s\right)
=\frac{0.1}{s+0.1}\varepsilon(s).$

The overall state-space representation is formulated into
(\ref{perturbedsystem}) as follows:%
\begin{align}
\dot{x}  &  =A\left(  \theta\right)  x+bu_{\xi}+d\nonumber\\
y  &  =c^{T}x \label{Two-Cart}%
\end{align}
where%
\begin{align*}
x  &  =\left[
\begin{array}
[c]{c}%
x_{1}\\
x_{2}\\
\dot{x}_{1}\\
\dot{x}_{2}%
\end{array}
\right]  ,A\left(  \theta\right)  =\left[
\begin{array}
[c]{cccc}%
0 & 0 & 1 & 0\\
0 & 0 & 0 & 1\\
-\frac{k_{1}}{m_{1}} & \frac{k_{1}}{m_{1}} & -\frac{b_{1}}{m_{1}} &
\frac{b_{1}}{m_{1}}\\
\frac{k_{1}}{m_{2}} & -\frac{k_{1}+k_{2}}{m_{2}} &
\frac{b_{2}}{m_{2}} &
-\frac{b_{1}+b_{2}}{m_{2}}%
\end{array}
\right]  ,d=\left[
\begin{array}
[c]{c}%
0\\
0\\
0\\
\frac{1}{m_{2}}%
\end{array}
\right]  \zeta,\\
b  &  =\left[
\begin{array}
[c]{c}%
0\\
0\\
\frac{1}{m_{1}}\\
0
\end{array}
\right]  ,c=\left[
\begin{array}
[c]{c}%
0\\
1\\
0\\
0
\end{array}
\right]  ,\theta=\left[
\begin{array}
[c]{cccccc}%
m_{1} & m_{2} & k_{1} & k_{2} & b_{1} & b_{2}%
\end{array}
\right]  .
\end{align*}
The objective here is to design a tracking controller $u$ such that
$y\rightarrow r$ with a good tracking accuracy. Since the dampers
will always consume the energy untile the two carts are at rest, it
can be concluded that the two-cart system (a physical system) is
stable for any $\theta$ in the real world. This implies that, for
any given $\theta\in\Omega,$ there exist
positive definite matrices $P\in%
\mathbb{R}
^{n\times n}$ and $Q\in%
\mathbb{R}
^{n\times n}$ such that $PA\left(  \theta\right)  +A^{T}\left(
\theta\right)
P\leq-Q,\forall x\in%
\mathbb{R}
^{n},$ where $\Omega$ represents the set that any $\theta$ in the
real world. \textit{Assumption 1} is satisfied. Let $C\left(
s\right)  =\frac{1}{2s+1}.$ Then \textit{Assumption 2 }is satisfied
with $\varepsilon_{H}=0.12\ $and $\varepsilon_{\tau}=1.$

According to input redefinition above, the two-cart system
(\ref{Two-Cart})
can be rewritten as follows:%
\begin{align*}
\dot{x}  &  =A\left(  \theta\right)  x+bu+\left(  d+\xi\right) \\
y  &  =c^{T}x
\end{align*}
where $\sup_{t\geq0}\left\vert \xi\left(  t\right)  \right\vert
\leq0.17a$ and
$u$ is generated by%
\begin{align*}
\dot{z}  &  =-0.5z+0.5\sigma_{a}\left(  v\right)  ,\\
u  &  =z.
\end{align*}
Then, according to (\ref{equ1_equivalent}), the uncertainty-free
transformed
system of (\ref{Two-Cart}) is%
\begin{align}
\dot{x}_{new}  &  =A(\hat{\theta})x_{new}+bu,x_{new}\left(  0\right)
=0\nonumber\\
y  &  =c^{T}x_{new}+d_{new}. \label{TwoCart_equa}%
\end{align}
where $d_{new}$ $=$ $c^{T}e^{A\left(  \theta\right)  t}x_{0}$ $+%
{\displaystyle\int\nolimits_{0}^{t}}
c^{T}e^{A\left(  \theta\right)  \left(  t-s\right)  }b[d\left(
s\right) +\xi\left(  s\right)  +A\left(  \theta\right)
x_{new}\left(  s\right) -A(\hat{\theta})x_{new}\left(  s\right)
]ds.$ According to (\ref{equ1_Pri}),
the primary system of (\ref{TwoCart_equa}) can be rewritten as follows:%
\begin{align}
\dot{x}_{p}  &  =A(\hat{\theta})x_{p}+bu_{p},x_{p}\left(  0\right)
=0\nonumber\\
y_{p}  &  =c^{T}x_{p}+d_{new}. \label{primaryTwoCart}%
\end{align}
The transfer function from $u_{p}$ to $y_{p}$ in
(\ref{primaryTwoCart}) is $G_{yu}\left(  s\right)  ,$ which is a
minimum phase. Thus, $u_{p}$ can be designed as $u_{p}\left(
s\right)  =G_{yu}^{-1}\left(  s\right)  \left( r-d_{new}\right)
\left(  s\right)  ,$ which can drive $y_{p}\rightarrow r.$
The \textit{Problem 1} is sloved. Furthermore, according to (\ref{easycontrol}%
), redefined input $v$ is designed as $v^{s}\left(  t,u_{p}\right)  =%
\mathcal{L}%
^{-1}\left(  C^{-1}\left(  s\right)  G_{yu}^{-1}\left(  s\right)
\left( r-d_{new}\right)  \left(  s\right)  \right)  .$ To realize
the control,
$v^{s}\left(  t,u_{p}\right)  $ is approximated to be%
\begin{equation}
v^{s}\left(  t,u_{p}\right)  =%
\mathcal{L}%
^{-1}\left(  Q\left(  s\right)  C^{-1}\left(  s\right)
G_{yu}^{-1}\left( s\right)  \left(  r-d_{new}\right)  \left(
s\right)  \right)  .
\label{approximated}%
\end{equation}
where $Q\left(  s\right)  $ is a fifth-order low-pass filter to make
the compensator physically realizable (the order of denominator is
greater than or equal to that of numerator). For simplicity,
$Q\left(  s\right)  $ is chosen to be $Q\left(  s\right)
=\frac{1}{\prod\nolimits_{k=1}^{k=5}\left(  \frac {1}{10k}s+1\right)
}$ here. The \textit{Problem 2} is sloved. Therefore, according to
(\ref{mainobsever1})-(\ref{maincontroller1}), the controller for
the two-cart system is designed as follows:%
\begin{align}
\dot{\hat{x}}_{new}  &
=A(\hat{\theta})\hat{x}_{new}+bu,\hat{x}_{new}\left(
0\right)  =0,\hat{d}_{new}=y-c^{T}\hat{x}_{new}\nonumber\\
\dot{z}  &  =-0.5z+0.5\sigma_{a}\left(  v^{s}\left(  t,u_{p}\right)
\right)
,u=z \label{approximatedcont}%
\end{align}
where $v^{s}\left(  t,u_{p}\right)  $ is given by
(\ref{approximated}) and $a\ $is chosen to be $1$ here.

To shown the effectiveness, the proposed controller
(\ref{approximatedcont})
is applied to three cases:%
\[%
\begin{array}
[c]{l}%
\text{Case 1: }\hat{\theta}=\theta\\
\text{Case 2: }\hat{\theta}=[%
\begin{array}
[c]{cccccc}%
1 & 1 & 1 & 0.9 & 1.5 & 1
\end{array}
]^{T}\\
\text{Case 3: }\theta=\hat{\theta}=[%
\begin{array}
[c]{cccccc}%
1 & 1 & 1 & 0.9 & 1.5 & 1
\end{array}
]^{T}.
\end{array}
\]
Case 1 implies the parameters are known exactly. Case 2 implies the
parameters are unknown. While, Case 3 implies the parameters are
changed to be a specified one. The simulations are shown in Figs.
5-7. The proposed controller achieves a good tracking accuracy.
Moreover, it is seen that the response in Cases 2-3 is faster than
that in Case 1. And, the tracking accuracy in Cases 1,3 is better
than that in Case 2. So, Case 2 is a tradeoff between Case 1 and
Case 3.

\textbf{Remark 8}. The simulations show that the proposed controller
can handle the case that the estimate parameters are different from
the true parameters. Moreover, the response is similar to that of
the model with the estimate parameters. This implies that the
proposed controller in fact achieves the results similar to model
reference adaptive control. However, unlike model reference adaptive
control, unknown parameters are not estimated and changed directly.

\textbf{Remark 9}. If the considered system is parameterized but
there exist many uncertain parameters, then an adaptive control
often needs corresponding number of estimators, i.e., corresponding
number of integrators. This will cause parameters converging to true
values with very slow rate or cannot converge to true values if
without persistent excitation. Whereas, in the proposed control,
five uncertain parameters and disturbance are lumped into the
disturbance $d_{new},$ which can be estimated directly.

\section{Conclusions}

Output tracking control for a class of uncertain systems subject to
unmodeled dynamics and time delay at input is considered. Our main
contribution lies on the presentation of a new decomposition scheme,
named additive state decomposition, which not only transforms the
uncertain system to be an uncertainty-free system but also
simplifies the controller design. The proposed control scheme is
with the following two salient features. (i) The proposed control
scheme can handle mismatching\textbf{ }uncertainties and mismatching
disturbance. Moreover, it can achieve a good tracking performance
without exact parameters. (ii) The proposed control scheme has
considered many uncertainties. In the presence of these
uncertainties, the closed-loop system is still stable when
incorporating the proposed controller. Three benchmark examples are
given to show the effectiveness of the proposed control scheme.

\begin{figure}[h]
\begin{center}
\includegraphics[
scale=0.7 ]{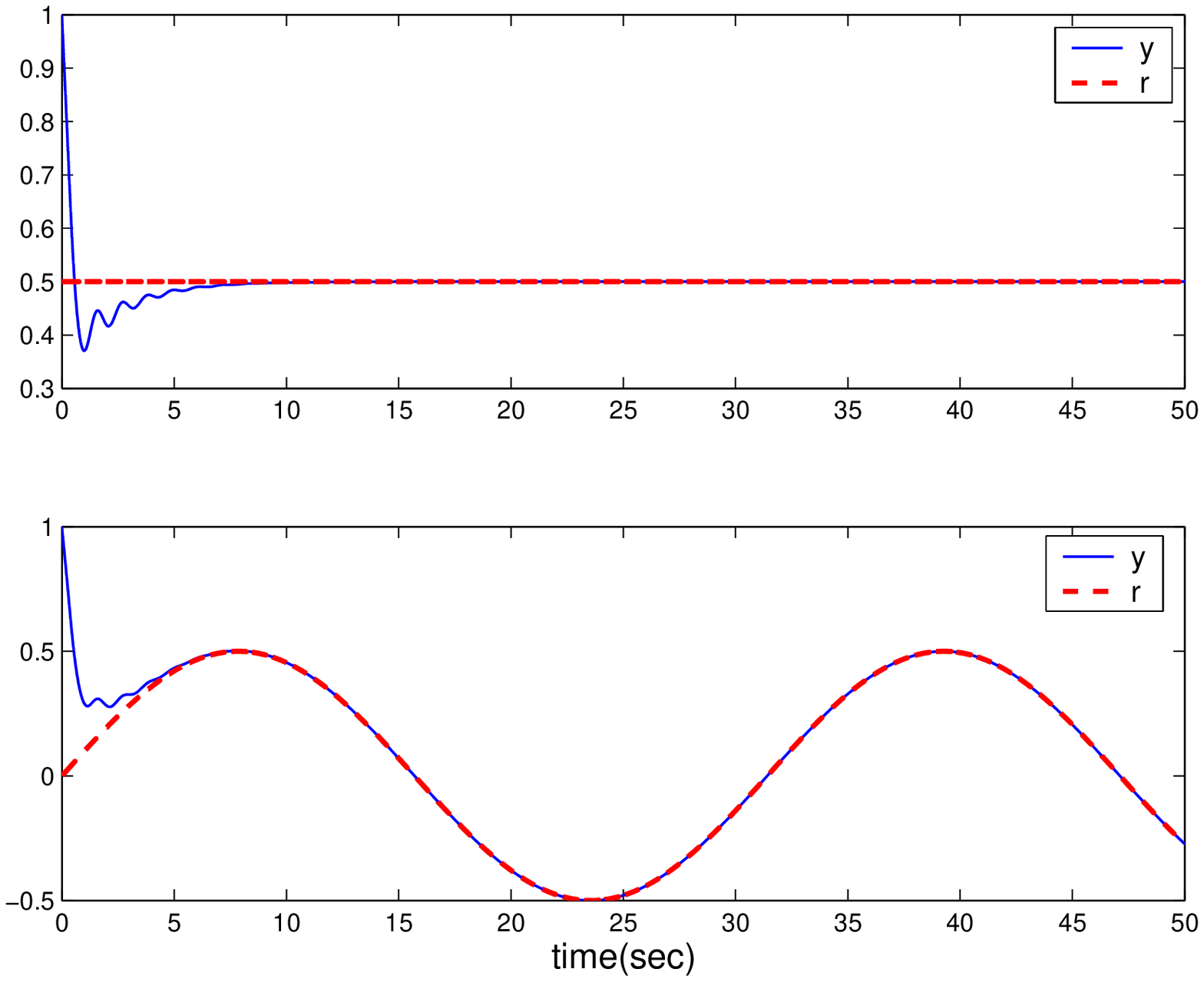}
\end{center}
\caption{Output of the Rohrs' example system}%
\end{figure}\begin{figure}[h]
\begin{center}
\includegraphics[
scale=0.7 ]{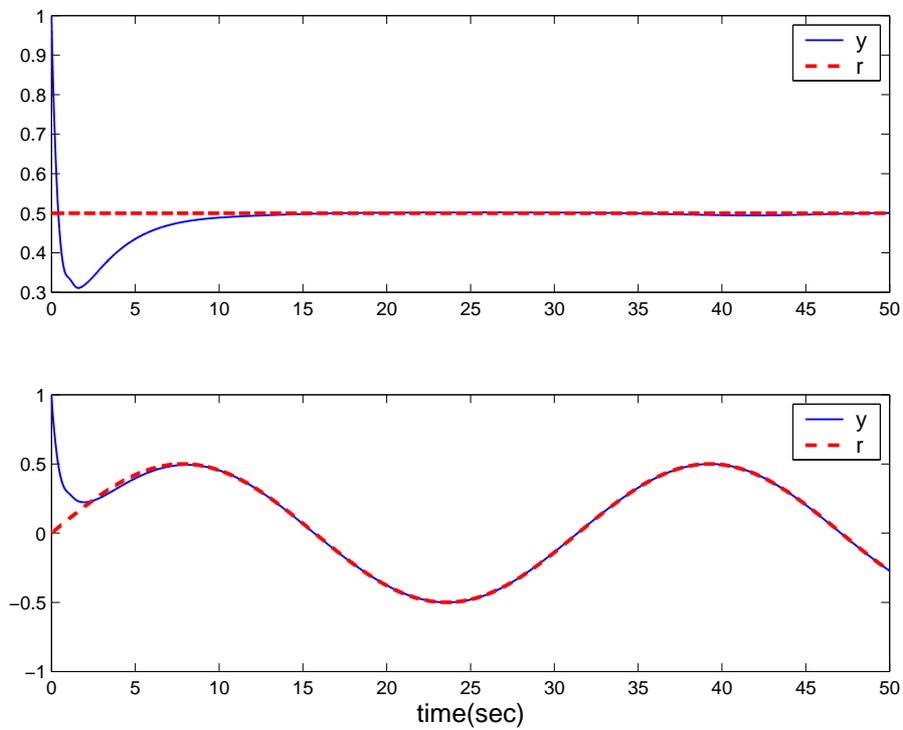}
\end{center}
\caption{Output of the nonlinear system}%
\end{figure}\begin{figure}[h]
\begin{center}
\includegraphics[
scale=0.9 ]{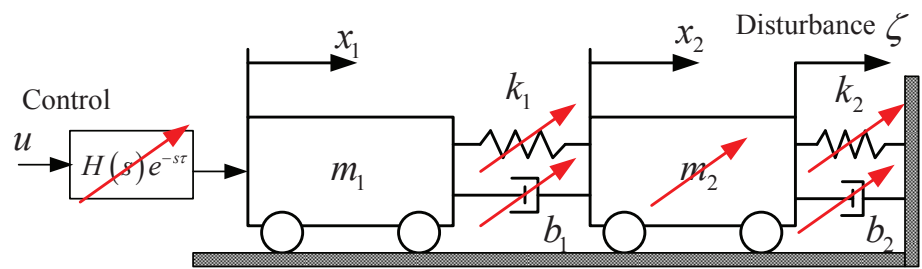}
\end{center}
\caption{The two-cart system}%
\end{figure}\begin{figure}[h]
\begin{center}
\includegraphics[
scale=0.7 ]{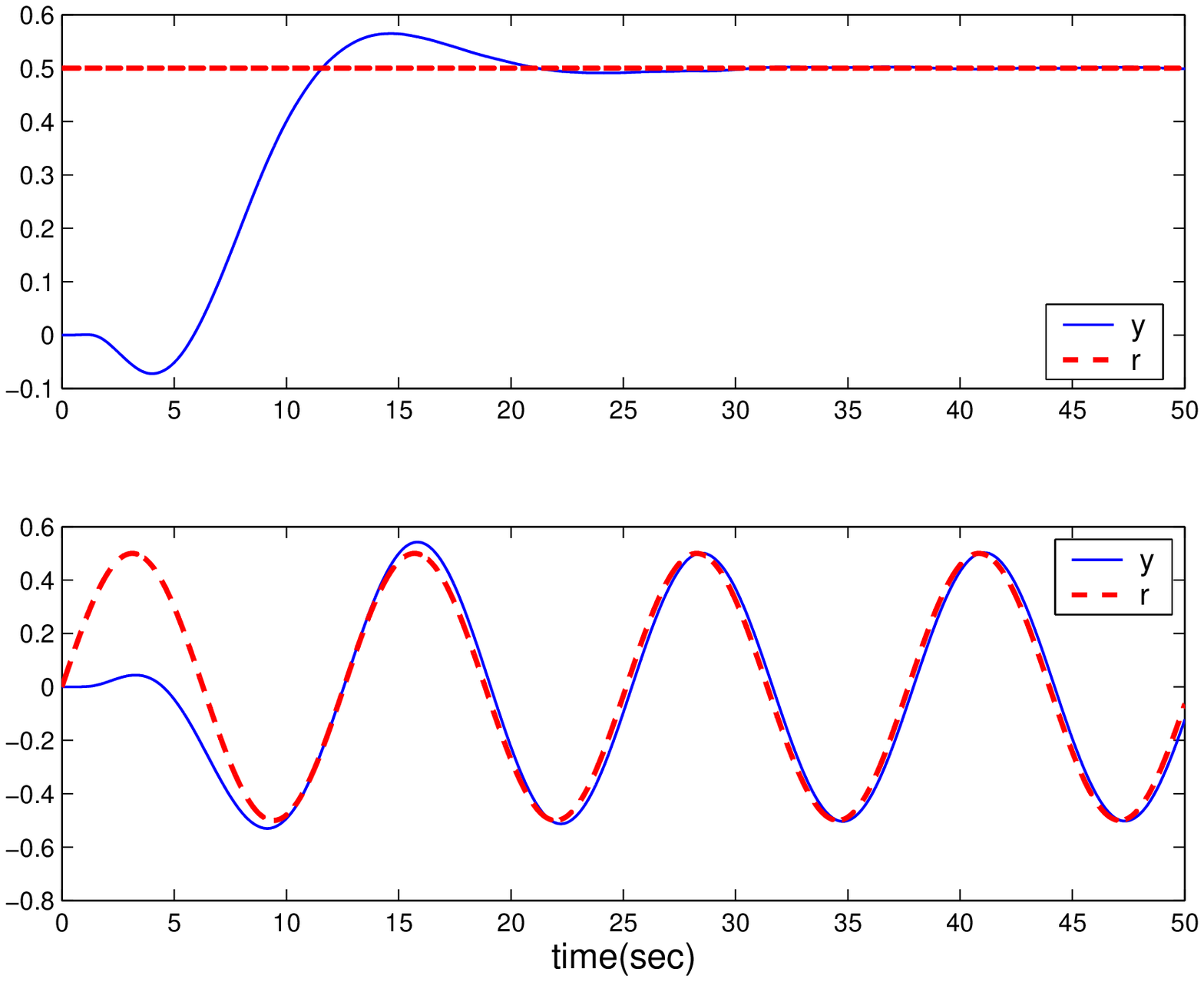}
\end{center}
\caption{Output of the two-cart system in Case 1}%
\end{figure}\begin{figure}[h]
\begin{center}
\includegraphics[
scale=0.7 ]{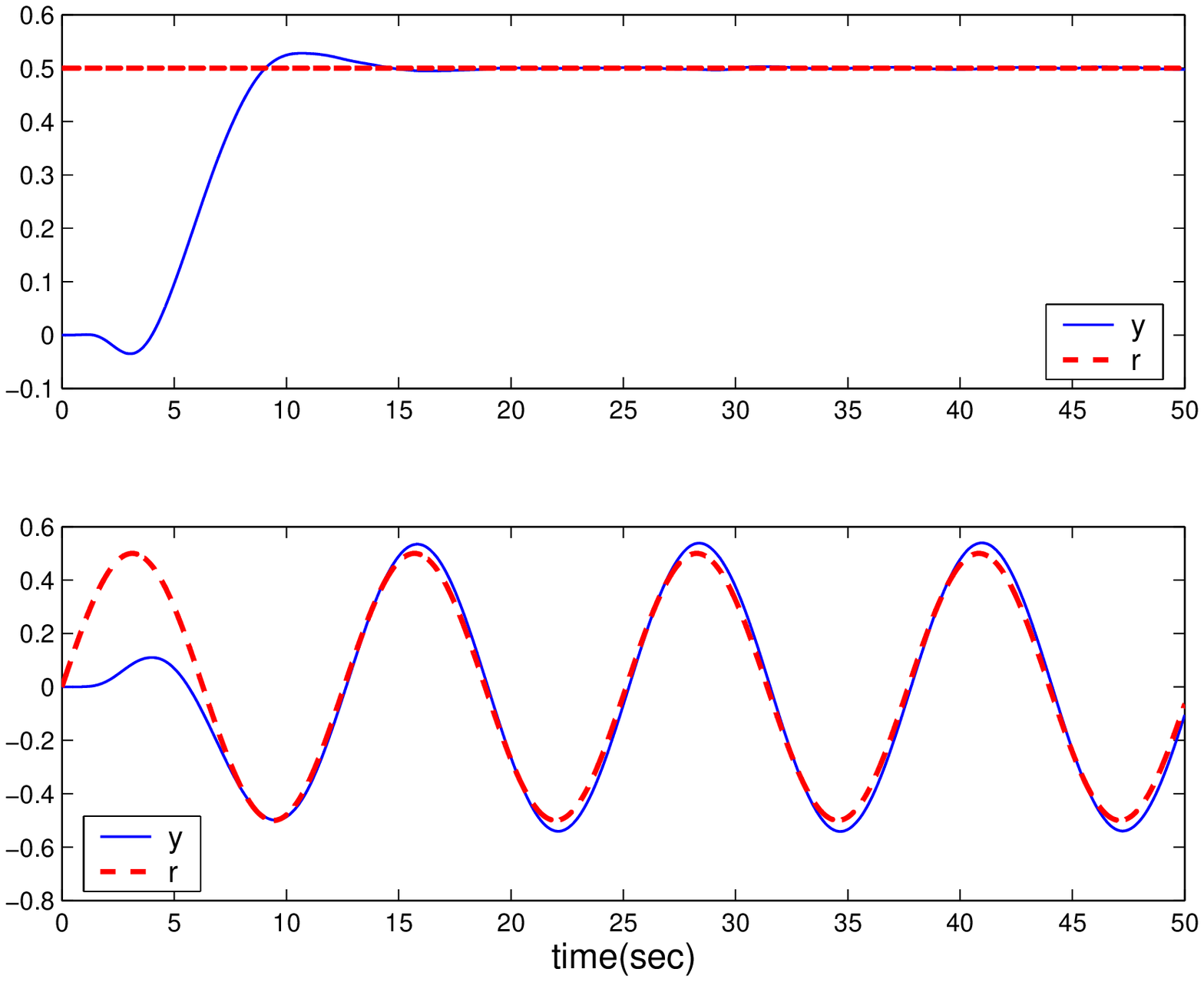}
\end{center}
\caption{Output of the two-cart system in Case 2}%
\end{figure}\begin{figure}[h]
\begin{center}
\includegraphics[
scale=0.7 ]{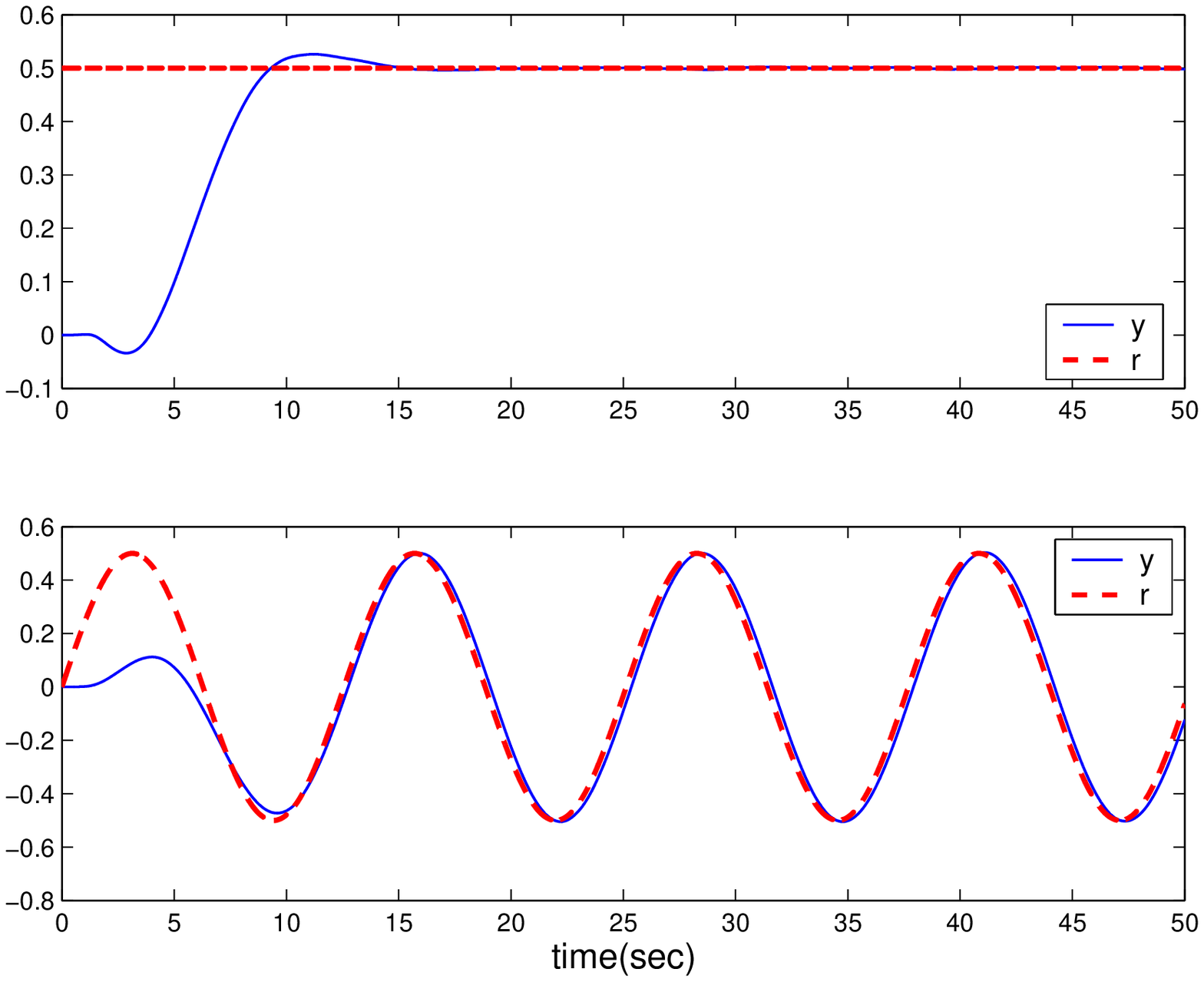}
\end{center}
\caption{Output of the two-cart system in Case 3}%
\end{figure}

\end{document}